\documentclass[aps,prl,a4paper,preprintnumbers,amsmath,amssymb,superscriptaddress,twocolumn,floatfix,showpacs,nofootinbib]{revtex4}
\usepackage{amssymb}
\usepackage{float,graphicx}

\begin{document}

\title{A Realistic Cosmology Without Cold Dark Matter}
\author{Baojiu~Li}
\email[Email address: ]{b.li@damtp.cam.ac.uk} \affiliation{DAMTP,
Centre for Mathematical Sciences, University of Cambridge,
Cambridge CB3 0WA, UK}
\author{Hongsheng~Zhao}
\email[Email address: ]{hz4@st-andrews.ac.uk} \affiliation{SUPA,
University of St Andrews, North Haugh, Fife, KY16 9SS, UK}
\affiliation{The Observatory, Leiden University, Niels Bohrweg 2,
Leiden, The Netherlands }
\date{\today}

\begin{abstract}
We propose a new framework unifying cold dark matter (CDM) and
Modified Newtonian Dynamics (MOND) to solve their respective
problems on galactic scales and large scale structure formation.
In our framework the dark matter clusters on large scales but not
on galactic scales. This \emph{environment dependence} of the dark
matter behaviors is controlled by a vector field, which also
produces the MOND effects in galaxies. We find that in this
framework only a \emph{single} mass scale needs to be introduced
to produce the phenomena of CDM, MOND and also dark energy.
\end{abstract}


\maketitle

\emph{Introduction}: The observed universe appears not to be
purely made of standard model particles and Einstein gravity. We
are yet to identify the physics of the missing constituent(s).
Data on galactic and larger scales are often used to argue for the
MOND and the CDM frameworks respectively. Linear growth of large
scale structure in the early universe, such as the Cosmic
Microwave Background (CMB), favors the CDM idea, and nonlinear
structures on the scale of galaxy clusters agree very well with
numerical CDM simulations. On the other hand, on smaller scales,
Milgrom's MOND formula \cite{Milgrom1983} captures the tight
correlation of the observed baryonic mass distribution within a
spiral galaxy vs.~the observed gravitational acceleration at each
radii of that galaxy. This applies to galaxies with a wide range
of scales, formation histories and environments, from dwarf to
elliptical galaxies \cite{Milgrom2007a, Milgrom2007b, Famaey2005,
Sanders2007, Angus2008a}. The amazing accuracy of this relation
and the fact that it predicts correct Tully-Fisher relation even
for tidal dwarf galaxies \cite{Gentile2007} motivate the
non-covariant MOND theory \cite{Bekenstein1984} and a class of
covariant theories \cite{Bekenstein2004, Sanders2005, Zhao2007,
Skordis2008, Zlosnik2007, Blanchet2008, Bruneton2008}, to
eliminate the need for the CDM particles.

However, both CDM and covariant MOND have their own problems. So
far the most challenging difficulty for covariant MOND theories is
in producing early growths of large scale structure and fitting
the CMB data \cite{Skordis2006, Dodelson2006, Zlosnik2008} on
which CDM works very well. Also, massive neutrinos seem to be
indispensable even in covariant MOND to explain the lensing data
in galaxy clusters \cite{Angus2007, Tian2008}. In comparison,
thanks to its simplicity, the CDM framework enjoys many tools for
sophisticated numerical simulations, yet the properties of
galaxies in these simulations are not in good agreement with
observations. The overproduction of dark structures in small
scales is well-known as the substructure problem and the cusp
problem. A common assumption to solve this problem is that CDM
particle is ballistic and will alway follows the same geodesic
equation as a collisionless star would.

Indeed, CDM and MOND are both complementary and mutually
exclusive: if both exist in galaxies, then obviously new problems
will arise. A natural way out of this dilemma is to have a "CDM"
which is no longer cold in the environments where MOND dominates,
\emph{e.g.}, letting it develop a nonzero pressure or have a much
smaller mass (for particles) there. This environment dependence
may be controlled by a scalar field, but in this case the
different dark matter behaviours in different regimes (galactic,
cluster and cosmological) indicate that the coupling between the
scalar field and the dark matter must be fine-tuned (if it is
possible anyway), because of the dynamical nature of the scalar
field. In order to see why, note that the MONDian behaviour is
only expected where the Newtonian acceleration $|\nabla\Phi|$ is
smaller than the MOND parameter $a_{0}$ (on galactic scales), but
not in the solar system (where $|\nabla\Phi|\gg a_{0}$), nor on
cosmological scales in most of the cosmic history (where $cH\gg
a_{0}$). These suggest that we should use \emph{both}
$|\nabla\Phi|/a_{0}$ \emph{and} $cH/a_{0}$ as the criteria about
the environment dependence of the dark matter behaviours.
A scalar field dark matter faces not only the challenge to
reproduce MOND when $|\nabla\Phi|\lesssim a_{0}$
\cite{Bruneton2007}, but also to follow \emph{both} $\nabla\Phi$ \emph{and} $cH$ through a
correct dynamical evolution, because
there are no inherent characteristic quantities which mimic
$\nabla\Phi$ or $cH$ in normal scalar field models.

\begin{table}[tbp]
\caption{The behaviours of the $c_{i}$ terms in different relevant
regimes ($+$ means there is effect and $-$ means no effect). The
static limit of the terms actually depend on spatial configuration
of the vector field, but this can be consistently regarded as of
higher order.} \label{tab:table1}
\begin{ruledtabular}
\begin{tabular}{ccccc}
Terms & $c_{1}$ & $c_{2}$ & $c_{3}$ & $c_{4}$ \\
\hline Cosmological background & $+$ &
$+$ & $+$ & $-$ \\
Cosmological perturbation & $+$ & $+$ & $+$ & $+$ \\
Static limit & $+$ & $-$ & $-$ & $+$ \\
\end{tabular}
\end{ruledtabular}
\end{table}

In most attempts to construct relativistic MOND, a vector field is
used, which can easily overcome the second challenge faced by
scalar fields. Furthermore, from previous studies of time-like
unit-norm vector fields (the \AE ther field $\AE_{a}$
\cite{Jacobson2001}), we know that there are four possible kinetic
terms for $\AE_{a}$, which have different properties in different
regimes. If we write these kinetic terms as $\mathcal{K}\equiv
K^{ab}_{\ \ cd}\nabla_{a}\AE^{c}\nabla_{b}\AE^{d}$ with $K^{ab}_{\
\ cd} = c_{1}g^{ab}g_{cd} + c_{2}\delta^{a}_{c}\delta^{b}_{d} +
c_{3}\delta^{a}_{d}\delta^{b}_{c} + c_{4}\AE^{a}\AE^{b}g_{cd}$, in
which $c_{i}$'s are dimensionless constants, then
Table~\ref{tab:table1} briefly summarizes the behaviours of these
terms in different regimes. More specifically, in the static limit
the $c_{1}$ and $c_{4}$ terms are $\propto(\nabla\Phi)^{2}$ while
in background the $c_{1,2,3}$ terms are $\propto (cH)^{2}$ (see
\cite{Zhao2007} for an earlier discussion about this point).

These facts suggest that the \AE ther field $\AE_{a}$ should be a
natural candidate to control the environment dependence of the
dark matter's behaviour, most probably through a coupling to the
latter. In this work we shall give a simple example to illustrate
this principle. The idea is very straightforward: let the
behaviour of the dark matter depend on $\mathcal{K}/a_{0}^{2}$,
which could be very large on cosmological scales in most of the
cosmic history ($\mathcal{K}\sim c^{2}H^{2}\gg a_{0}^{2}$) and in
the solar system ($\mathcal{K}\sim|\nabla\Phi|^{2}\gg a_{0}^{2}$),
but becomes of order unity or less on galactic scales
$\mathcal{K}\sim|\nabla\Phi|^{2}\lesssim a^{2}_{0}$. In the first
case the dark matter has (nearly) zero pressure and zero sound
speed, and thus is actually cold, while in the latter situations
it acquires a nonzero pressure and nonzero sound speed, and thus
no longer clusters; instead then, the $\AE_{a}$ field will produce
the MOND effect.

\footnotetext[1]{Note that in Eq.~(\ref{eq:EMT_1}) the
$\mathcal{K}$ in $g(X,\mathcal{K})$ can be different from that in
$V(\varphi,\mathcal{K})$ (\emph{e.g.}, having different
$c_{i}$'s). Consequently, if for example the $\mathcal{K}$ in
$V(\varphi,\mathcal{K})$ has no $c_{4}$ term, then
$c_{4}V_{\mathcal{K}}$ is zero, but this does not necessarily mean
that $c_{4}g_{\mathcal{K}}$ is also zero and vice versa.}

\emph{The Model}: A distinction between our model and previous
ones \cite{Bekenstein2004, Sanders2005, Zhao2007, Skordis2008,
Zlosnik2008} is that we are not using $\AE_{a}$ to grow the large
scale structure, which proves difficult. Instead, we introduce a
dark matter $\varphi$ which is \emph{controlled by} $\AE_{a}$
through a coupling. For illustration purpose, we take $\varphi$ as
a k-essence field \cite{kessence}, though our principle can be
applied much more generally (see a discussion below). We start
from the following Lagrangian density
\begin{eqnarray} \label{eq:Lagrangian}
\mathcal{L} &=& -g(\varphi,X,\mathcal{K}) + V(\varphi,\mathcal{K})
+ \lambda\left(\AE^{a}\AE_{a}-1\right)
\end{eqnarray}
in which $g, V$ are arbitrary functions, $\varphi$ is a
dimensionless scalar field,
$X\equiv\frac{1}{2}\nabla^{a}\varphi\nabla_{a}\varphi$ and
$\mathcal{K}, \AE_{a}$ are defined above; $\lambda$ is a Lagrange
multiplier ensuring that $\AE_{a}$ has unit norm.  For $g(\varphi,X,\mathcal{K})$, we
use a generic power-law in $X$ with the power
dependent on $\mathcal{K}$:
\begin{eqnarray}\label{eq:1_g}
g(\varphi,X,\mathcal{K})\ =\ g(X,\mathcal{K})\ =\
2w\left(\frac{X}{a_0^2}\right)^{\frac{w+1}{2w}}a_0^2
\end{eqnarray}
in which the normalization $a_0^2$ is always justified because
$\varphi$ (hence $X$) could be rescaled;
$w=w\left(\frac{\mathcal{K}_{3}}{a^{2}_{0}}\right)$, which has the
meaning of the dark matter equation of state, is a free function
introduced to produce the desired effects, and
$\mathcal{K}_{3}\equiv
c_{3}\delta^{a}_{d}\delta^{b}_{c}\nabla_{a}\AE^{c}\nabla_{b}\AE^{d}$
\emph{only} has the $c_{3}$ term.  Here $c_3$ is a constant, and we adopt $\frac{1}{w} =
3+\left(\frac{\mathcal{K}}{a_{0}^{2}}\right)^{n}$, where $n=1$ for
illustration.
We further
choose $V(\varphi,\mathcal{K})=V\left(\mathcal{K}_4\right)$ where
$\mathcal{K}_{4}\equiv
c_{4}(w)\AE^{a}\AE^{b}g_{cd}\nabla_{a}\AE^{c}\nabla_{b}\AE^{d}$ has
\emph{only} the $c_{4}$ term; here $c_4$ can be a function of $w$, but for
the simplicity of arguments, we shall first treat $c_4$ as a constant first.
Note that our choices of the $c_{3}$ and $c_{4}$
terms above are designed so that we have a clean separation
between the dark matter behaviours in cosmological background
($g$) and in the static limit ($V$). This is achieved because the
$c_{3}$ term only has effects on the former while the $c_{4}$ term
only affects the latter (c.f.~Table~\ref{tab:table1}).

A variation with respect to the metric tensor gives the following
energy momentum tensor \footnotemark[1]
\begin{eqnarray}\label{eq:EMT_1}
&&8\pi GT_{ab}\nonumber\\ &=&
g_{X}\nabla_{a}\varphi\nabla_{b}\varphi -
g_{ab}\left[g(X,\mathcal{K})-V(\mathcal{K})\right]\nonumber\\
&& - 2\nabla_{c}\left[\left(V_{\mathcal{K}}-g_{\mathcal{K}}\right)
\left(\AE^{c}J_{(ab)}+\AE_{(a}J^{c}_{\ b)}-\AE_{(a}J_{b)}^{\
c}\right)\right]\nonumber\\
&& + 2\nabla_{c}\left[\left(V_{\mathcal{K}}-g_{\mathcal{K}}\right)
J^{cd}\right]\AE_{d}\AE_{a}\AE_{b}\nonumber\\
&& - 2c_{4}V_{\mathcal{K}}
\left(\AE^{d}\nabla_{d}\AE_{c}\right)\left(\AE^{e}\nabla_{e}\AE^{c}\right)\AE_{a}\AE_{b}\nonumber\\
&& + 2c_{4}V_{\mathcal{K}}
\left(\AE^{c}\nabla_{c}\AE_{a}\right)\left(\AE^{d}\nabla_{d}\AE_{b}\right),
\end{eqnarray}
where we have defined $ J^{a}_{\ c}\equiv K^{ab}_{\ \
cd}\nabla_{b}\AE^{d}$ and
\begin{eqnarray}
g_{\mathcal{K}} &\equiv& \frac{\partial
g(X,\mathcal{K})}{\partial\mathcal{K}},\ \ V_{\mathcal{K}}\
\equiv\ \frac{\partial
V(\varphi,\mathcal{K})}{\partial\mathcal{K}}.
\end{eqnarray}
We could also derive the scalar and vector field equations of
motion, but they are not needed here.

The energy momentum tensor for the k-essence field is defined as
(see below for a discussion)
\begin{eqnarray}\label{eq:EMT_1_scalar}
8\pi GT_{ab}^{\varphi} &=& g_{X}\nabla_{a}\varphi\nabla_{b}\varphi
- g_{ab} g(X,\mathcal{K})
\end{eqnarray}
which resembles the energy momentum tensor of a perfect fluid
$T^{\varphi}_{ab} =
(\rho_{\varphi}+p_{\varphi})u_{a}u_{b}-p_{\varphi}g_{ab}$ with
$u_{a}=\frac{\nabla_{a}\varphi}{\sqrt{2X}}$ and energy density
$\rho_{\varphi}$ and pressure $p_{\varphi}$:
\begin{eqnarray}\label{eq:1_rho_phi}
8\pi G\rho_{\varphi} &=& 2Xg_{X}-g(X,\mathcal{K}),\\
\label{eq:1_p_phi} 8\pi Gp_{\varphi} &=& g(X,\mathcal{K}).
\end{eqnarray}
Substituting Eq.~(\ref{eq:1_g}) into Eqs.~(\ref{eq:1_rho_phi},
\ref{eq:1_p_phi}) it is easy to check
\begin{eqnarray}\label{eq:1_EOS}
w &=& \frac{p_{\varphi}}{\rho_{\varphi}}\ =\
\frac{1}{3+\frac{\mathcal{K}}{a_{0}^{2}}}.
\end{eqnarray}
So we see that when $\mathcal{K}\gg a_{0}^{2}$ this behaves as
dust while when $\mathcal{K}\ll a_{0}^{2}$ it behaves as
radiation. Furthermore, when it behaves as dust the sound speed
$c_{s}^{2}$ satisfies
\begin{eqnarray}
c^{2}_{s} &=& \frac{g_{X}}{2Xg_{XX}+g_{X}}\ =\
\frac{1}{3+\frac{\mathcal{K}}{a_{0}^{2}}}\ \rightarrow\ 0
\end{eqnarray}
so that it has the desired clustering property of CDM.

Remember that we have chosen the $\mathcal{K}$ in
$w\left(\frac{\mathcal{K}}{a_{0}^{2}}\right)$ as to have only a
$c_{3}$ term. It requires no fine tuning to choose $c_{3}$ so that
all through the cosmic history $\mathcal{K}\sim
3c_{3}H^{2}c^{2}\gg a^{2}_{0}$, since $cH_{0}\sim 6a_{0}$ and
$H>H_{0}$ at earlier times, as the result of which a choice of
$c_{3}\sim\mathcal{O}(10^{0}-10^{2})$ guarantees ${1\over w} \sim
\frac{\mathcal{K}}{a_{0}^{2}} \ge108c_{3}\gg1$ (if we use $n>1$ in
$w$ then $c_{3}$ could even be set to $1$). This indicates that in
our model the scalar field does behave like CDM in the background
expansion and the large scale structure formation, where there is
significant Hubble expansion.

At this stage one may be worried about the other terms in the
energy momentum tensor in Eq.~(\ref{eq:EMT_1}): are they large
enough to spoil the good CDM behaviours we have obtained so far?
We discuss how $V(\mathcal{K})$ and $V_{\mathcal{K}}$ are
negligible separately below. For the $g_{\mathcal{K}}$ terms, from
Eq.~(\ref{eq:1_g}) we get
\begin{eqnarray}
g_{\mathcal{K}} = \frac{8\pi
Gp_{\varphi}}{2a_{0}^{2}}\left[\log\frac{X}{a_0^2}-2w\right].
\end{eqnarray}
To see that $g_{\mathcal{K}}\ll1$, note that
$p_{\varphi}=w\rho_{\varphi}\sim
\rho_{\varphi}a_{0}^{2}/\mathcal{K}$ when $\mathcal{K} =
3c_{3}(cH)^2 \gg a_{0}^{2}$ and $8\pi G\rho_{\varphi}/3(cH)^2 \sim
O(1)$, so
\begin{eqnarray}\label{eq:1_factor1}
\frac{8\pi Gp_{\varphi}}{2a_{0}^{2}} &\sim& \frac{4\pi
G\rho_{\varphi}}{\mathcal{K}}\ \sim\ \frac{4\pi
G\rho_{\varphi}}{c^{2}H^{2}}\ \sim\ \mathcal{O}(1).
\end{eqnarray}
Meanwhile, the current fractional energy density of dark matter is
$0.2$, which means that $8\pi
G\rho_{\varphi0}\sim0.6(cH_{0})^{2}\sim20a^{2}_{0}$, so we have
$8\pi G\rho_{\varphi}=
2\left(\frac{X}{a_0^2}\right)^{\frac{w+1}{2w}}a_0^2 \sim
20a^{2}_{0}(1+z)^{3}$ in which $z$ is the redshift, or
$\left(\frac{X}{a_0^2}\right)^{\frac{w+1}{2w}} = B (1+z)^{3}$ with
$B\sim 10$. As a result $\log\frac{X}{a_{0}^{2}} =
\frac{2w}{w+1}[\log B + 3\log (1+z)]\sim\mathcal{O}(w)\ll1$ today;
$\log\left(\frac{X}{a_{0}^{2}}\right)/w$ increases with redshift
logarithmically at high redshifts, \emph{e.g.},
$\log\left(\frac{X}{a_{0}^{2}}\right)/w\sim\mathcal{O}(10^{2})$ at
$z\sim10^{10}$. But $w\propto(1+z)^{-3}$ and $(1+z)^{-4}$ in the
matter and radiation dominated eras so that indeed both $w$ and
$\log\left(\frac{X}{a_{0}^{2}}\right)$ decrease quickly with
redshift. This above analysis shows that
$g_{\mathcal{K}}\sim\mathcal{O}(10^{0}-10^{2})w\ll1$ in all the
cosmological epochs of interests, which is easy to understand
because $g\sim a_{0}^{2}$ is very small while
$\frac{\mathcal{K}}{a^{2}_{0}}$ is very large (this
order-of-magnitude estimate holds for general $n$s). The smallness
of $g_{\mathcal{K}}$ strongly suppresses the effects of the \AE
ther terms in Eq.~(\ref{eq:EMT_1}), making them negligible. In
fact, the $T^{\varphi}_{ab}$ in Eq.~(\ref{eq:EMT_1_scalar}) is not
conserved, but the smallness of $g_{\mathcal{K}}$ implies that the
energy exchange between $\varphi$ and $\AE_{a}$ is just
negligible. Then, as $\frac{p}{\rho}, \frac{\dot{p}}{\dot{\rho}},
\frac{\delta p}{\delta\rho}\sim \mathcal{O}(w)\ll1$, Eqs.~(35, 36)
of \cite{Li2008} show that the perturbation growth also mimics
that of CDM for reasonable parameters. Numerical results will be
reported in a forthcoming paper.

We next consider the cluster scales, where the observations are
not compatible with MOND alone but necessarily incurs a certain
amount of dark matter. An example is the bullet cluster, in which
the offset between the gas and dark matter distributions is hard
to be explained by MOND. These scales generally have not decoupled
from the background expansion, where according to our model the
dark matter is still cold. Thus this model has the potential to
explain the observations on cluster scales.

On galactic scales, the spacetime is more or less static, which
means that $\frac{\mathcal{K}_{3}}{a_{0}^{2}}$ is small enough to
make $w=c^{2}_{s} \rightarrow\frac{1}{3}$, so that the pressure
support is strong enough to prevent any further collapse of dark
matter. This eliminates the CDM in galaxy systems as we expected,
since otherwise CDM and MOND will coexist, spoiling MOND's good
fit with data. Now, with the scalar field dark matter not
clustering and the $g_{\mathcal{K}}$ (with only $c_{3}$ term)
having no effect in static weak field, it is the
r$\mathrm{\hat{o}}$le of the $V(\mathcal{K})$ (only $c_{4}$ term)
to produce the MOND effect. To do this, let us use the metric
$ds^{2}=(1+2\epsilon\Phi)dt^{2}-(1-2\epsilon\Psi)dx^{i}dx^{j}$ and
write $\AE^{a}=\delta^{a}_{0}+\epsilon\ae^{a}$ in which $\ae^{a}$
is the perturbation of $\AE^{a}$ and $\epsilon$ is a small
positive quantity. Then up to first order in $\epsilon$, it is
easy to derive that $G_{00} = -2\Phi^{,i}_{\ ,i} =
2\partial_{i}\partial_{i}\Phi$ where we have used the fact that
$\Phi=\Psi$ thanks to the absence of anisotropic stresses. For the
energy density of the fields [c.f.~the right hand side of
Eq.~(\ref{eq:EMT_1})], we already know that the first line as well
as all $g_{\mathcal{K}}$ terms have negligible or zero effects,
also it is easy to show that up to first order in $\epsilon$ the
last three lines all vanish, while the second line reduces to
$-2\nabla_{i}\left(c_{4}V_{\mathcal{K}}\Phi^{,i}\right)$.
Defining $\mu \equiv 1-c_{4}V_{\mathcal{K}}$
the Poisson equation now reads (actually there is also a
$V(\mathcal{K})$ on the right hand side of
Eq.~(\ref{eq:1_Poisson}), but this is like a cosmological constant
and will not cluster)
\begin{eqnarray}\label{eq:1_Poisson}
2\partial_{i}\left[\mu(x)\partial_{i}\Phi\right] &=& 8\pi
G\rho_{b}
\end{eqnarray}
where $\rho_{b}$ is the local baryon energy density and the
argument of $\mu(x)$ is $x \equiv
\left(\frac{\mathcal{K}}{a_{0}^{2}}\right)^{\frac{1}{2}} =
\frac{|\nabla\Phi|}{(-c_4)^{-1/2}a_0}$ where we have used $\mathcal{K}=
-c_{4}|\nabla\Phi|^{2}$ up to $\mathcal{O}(\epsilon^{2})$ in the
static limit. We could choose the form of $V(\mathcal{K})$ or
$\mu$ as in \cite{Zhao2007}
\begin{eqnarray}\label{eq:1_V}
1-\mu(x)=\left(1+{x \over 3}\right)^{-3} &=& \frac{
V\left(\mathcal{K}\right)}{V(0)\left(1+{x \over
3}\right)\left(1+{2x \over 3}\right)},\ \ \
\end{eqnarray}
where $V(0) = (-c_4)^{-1}(3a_0)^2$. Clearly the MOND limit
$\mu(x)\rightarrow x $ is recovered when $x \rightarrow 0$ if we
choose $c_4=-1$.  In general $V(\mathcal{K})$ serves as a
non-uniform dark energy potential \cite{Zhao2007}, whose local
minimum can recover the MOND equation, and the background value
behaves as a cosmological constant far away from galaxies $V(0) =
(3a_0)^2$. Such a $V(0)$ with $c_4=-1$ is however not enough to
account for the dark energy with $8\pi G\rho_{\mathrm{DE}}\sim
81a_{0}^{2}$, and we will come back to this point later.

In the solar system, again the k-essence field $\varphi$ does not
cluster, and the $c_{3}$ term in $g_{\mathcal{K}}$ has no effect.
But here the MOND effect and cosmological constant effect are both
suppressed because from Eq.~(\ref{eq:1_V}) $V(\mathcal{K})
\rightarrow 54x^{-1} a_0^2 \ll a^{2}_{0}$, and $\mu(x) \rightarrow
1-(1+x/3)^{-3} \rightarrow 1$ in the strong gravity regime in
which $x \gtrsim \mathcal{O}\left(10^{6}\right) \gg 1$. In fact
the Newtonian gravity and the PPN limits are recovered
\cite{Zhao2007}.

\emph{Discussion}: We want to point out that the model described
above is only a very simple one for the Lagrangian
Eq.~(\ref{eq:Lagrangian}). One can also, for example, use the
oscillation of a canonical scalar field around its potential
minimum to provide the dark matter, with the steepness of the
potential depending on $\mathcal{K}/a^{2}_{0}$. In a more
phenomenological way, we could simply postulate a coupling between
dark matter particles and the vector field (like the coupling with
a scalar field) as a result of which the dark matter particle mass
depends on $\frac{\mathcal{K}}{a_{0}^{2}}$. Furthermore, it is
also interesting to see if the parameter $a^{2}_{0}$ is indeed
determined dynamically. These possibilities will be considered in
details in forthcoming papers.

The interesting fact $a_{0}\sim cH_{0}$ suggests that there may be
some fundamental relations between MOND and dark energy. In fact,
there are many possible ways by which our model can be generalized
to include dark energy as well. One way is that at late times when
$H \sim H_{0}$ the dark matter decays into dark energy
(\emph{e.g.}, its equation of state $w$ becomes $-1$). The idea
here is to use the quantity $a_{0}$ to determine both the
transitions from CDM to MOND and from CDM to dark energy. A more
straightforward method is to have a cosmological constant in
$V(\mathcal{K})$: as is shown above, the MOND effect only depends
on $V_{\mathcal{K}}$ but not $V(0)$, and we can use the dark
energy density to fix $V(0)$ so that the combination of dark
energy and MOND completely determines $V(\mathcal{K})$. Another
interesting possibility is to note that in Eq.~(\ref{eq:1_V}) the
MOND effect requires $c_{4}=-1$ while dark energy requires
$c_{4}\sim-\frac{1}{9}$. This can be easily achieved, again using
our principle of environment dependence: let $c_{4}$ depend on
$\mathcal{K}_{3}$, for example with
$c_{4}=-\left(3-6w\right)^{-2}$. In this case $V(\mathcal{K})$
acts as an environment dependent cosmological constant, which
accounts for the cosmic acceleration in background cosmology
($w\rightarrow0$) and approaches zero in the solar system
($w\rightarrow1/3$). Note that dark energy and MOND are unified
with a single $V(\mathcal{K})$ in the latter two possibilities,
and we thus have a  full Lagrangian as
\begin{eqnarray}\label{eq:Lag_final}
\mathcal{L} &=& - 2w a_{0}^{2}
\left[\frac{X}{a_{0}^{2}}\right]^{1+w \over 2w} + V(\mathcal{K})
\end{eqnarray}
in which $V(\mathcal{K}) = (9a_0)^{2}\left(1-2 w\right)^2
(1+\frac{2x}{3})/(1+\frac{x}{3})^2$ and
$x=\sqrt{\mathcal{K}_{4}}/a_{0}$ for the third possibility.
Interestingly, $a_{0}\sim cH_{0}$ is a single mass scale
introduced for this model to relate CDM, MOND and dark energy
together. All the other parameters ($c_i$s, $n$) are dimensionless
and $\sim\mathcal{O}(1)$, and there are no fine-tunings of them:
the huge difference between the dark matter density
$\rho_{\varphi}$ at earlier times and the scale ${a_0^2 \over 8\pi
G}$ comes as a generic result of the dynamical evolution of the
vector field. In this sense t\emph{he vector field acts as a
leverage, making the tiny mass scale $a_{0}$ capable to
characterize the large energy density of dark matter}. Meanwhile,
this could also shed further light on the dark energy coincidence
problem, since the dark energy dominance begins at the time when
galaxies have formed (and we observers come into existence), both
characterized by our fundamental mass scale $a_{0}$.

\emph{Summary}: In this work we have tried to tackle the problem
of how to unify CDM and MOND in a consistent way. The idea is to
give the dark matter an environment dependence, making it behave
like CDM on large scales, while reproducing the MOND (Newtonian
dynamics) in the static and weak (strong) field limits
respectively. Although the idea of an environment dependence is
not new, it is novel to use the vector ($\AE_{a}$) field as the
switch. We show how the particular properties of the vector field
make it very effective for this purpose. Our model provides a
general framework which can potentially solve the problems of CDM
on galactic scales and of MOND on larger scales. It could also be
generalized to include dark energy in a way such that all the
phenomena of CDM, MOND and dark energy are related to one
parameter $a_{0}$, which is the single mass scale introduced in
our model. Both fields in Eq.~(\ref{eq:Lag_final}), likely
effective, should provide insights to people seeking the
fundamental fields in particle physics theories.

\

\begin{acknowledgments}
\emph{Acknowledgments}: B.~Li and H. Zhao are indebted to the
HPC-Europa Transnational Access Visit programme for its support
and the Lorentz Center and Leiden Observatory for hospitality when
this work is undertaken.
\end{acknowledgments}

\bigskip


\end{document}